\documentclass[times,twocolumn,final]{elsarticle}
\usepackage{cag}
\usepackage{framed,multirow}

\usepackage{amssymb}
\usepackage{latexsym}

\usepackage{url}

\usepackage{xcolor}
\definecolor{newcolor}{rgb}{.8,.349,.1}

\usepackage{hyperref}

\usepackage[switch,pagewise]{lineno} %Required by command \linenumbers below

\journal{Computers \& Graphics}

\begin{document}

\verso{Preprint Submitted for review}

\begin{frontmatter}

\title{Underwater enhancement based on a self-learning strategy and attention mechanism for high-intensity regions}%
\tnotetext[tnote1]{}

\author[1]{Claudio D. \snm{Mello Jr.}\corref{cor1}}
\cortext[cor1]{Corresponding author: 
  Tel.: +55-053-3293-5106;  
  fax: +55-053-3293-5105;}
\emailauthor{claudio.mello@furg.br}{Claudio D. Mello Jr.}
%\ead{example@email.com}
    
\author[1]{Bryan U. \snm{Moreira}%\fnref{fn1}
}
%\fntext[fn1]{Footnote 1.}  

\author[1]{Paulo J. D. O.   \snm{Evald}%\fnref{fn1}
}
%\fntext[fn1]{Footnote 1.} 

\author[1]{Paulo J. L. \snm{Drews Jr.}%\fnref{fn1}
}
%\fntext[fn1]{Footnote 1.} 

\author[1]{Silvia S. C. \snm{Botelho}%\fnref{fn1}
}
%\fntext[fn1]{Footnote 1.} 

\address[1]{Center for Computer Science, Federal University of Rio Grande\\ Av. Italia - km8, Rio Grande, RS, 96203-900, Brazil}

%\received{1 February 2017}
\received{\today}
%%%% Do not use the below for submitted manuscripts
%\finalform{28 March 2017}
%\accepted{2 April 2017}
%\availableonline{15 May 2017}
%\communicated{S. Sarkar}

\begin{abstract}
%%%
Images acquired during underwater activities suffer from environmental properties of the water, such as turbidity and light attenuation. These phenomena cause color distortion, blurring, and contrast reduction. In addition, irregular ambient light distribution causes color channel unbalance and regions with high-intensity pixels. Recent works related to underwater image enhancement, and based on deep learning approaches, tackle the lack of paired datasets generating synthetic ground-truth. In this paper, we present a self-supervised learning methodology for underwater image enhancement based on deep learning that requires no paired datasets. The proposed method estimates the degradation present in underwater images. Besides, an autoencoder reconstructs this image, and its output image is degraded using the estimated degradation information. Therefore, the strategy replaces the output image with the degraded version in the loss function during the training phase. This procedure \textit{misleads} the neural network that learns to compensate the additional degradation. As a result, the reconstructed image is an enhanced version of the input image. Also, the algorithm presents an attention module to reduce high-intensity areas generated in enhanced images by color channel unbalances and outlier regions. Furthermore, the proposed methodology requires no ground-truth. Besides, only real underwater images were used to train the neural network, and the results indicate the effectiveness of the method in terms of color preservation, color cast reduction, and contrast improvement.
\end{abstract}


\end{frontmatter}

%\linenumbers

\section{Introduction}\label{sec1}

Modern underwater (UW) activities such as monitoring, inspection, maintenance, archaeology, and environmental research, involve the acquisition of video footage and images of objects, fauna and flora \citep{donaldson2020,drews2016a}. The quality of perception of these objects in the scene depends on physical properties of water, ambient light, and depth \citep{santos2020}. Turbidity is defined by the particles of organic and inorganic materials in suspension in water. They cause scattering and absorption of light rays causing blurred, and dim images \citep{ancuti2012,drews2016}. The turbidity and distance of the objects from the camera define the intensity of these phenomena. The depth of the scene affects the color perception, and the light components with longer wavelength are the first to be attenuated \citep{han2018}. The ambient light interacts with the particles in suspension in the water, and increases the turbidity perception in the image acquisition. Also, the nature of the material in suspension defines a brownish, greenish or blueish tonality or the \textit{color cast} of the water \citep{pan2019}. %\par

The UW image restoration methods describe the image using the Image Formation Model (IFM) \citep{akkaynak2018,drews2016}. The IFM represents the scene captured by the camera at a given depth, considering water attenuation and ambient light. The UW light attenuation depends on geographic and environmental conditions of the water \citep{berman2019}, and produces diversity in the image scenes.
The variability of turbidity, ambient light, and color cast represent a relevant challenge to UW image improvement tasks, given that the attenuation coefficients, distance of the objects from the camera and background light are unknown in most of the real image acquisition situations.\par

Image enhancement proposals have explored the improvement of the contrast and color, focusing on pixel intensity re-distribution via spatial-domain \citep{ancuti2018}, transform-domain \citep{vasamsetti2017} or Convolutional Neural Network (CNN)-based image enhancement \citep{li2021uda}. Recently, mixed approaches involving enhancement/restoration (dehazing) and IFM-Deep Learning (DL) based methods were presented in \citep{dudhane2020} and \citep{yufei2020}. Also, strategies based on DL and multi-color spaces with adaptive frameworks were used in \citep{cli2021} and \citep{wang2021}. \par

The improvement of the UW images based on DL is a difficult task, because the real reference image (\textit{real} ground-truth) is, normally, unavailable. Therefore, it is common the authors to resort to synthesized reference images. The strategies used for synthesizing ground-truth images are physical model-based techniques \citep{dudhane2020} or more complex structures performing style transfer \citep{cho2020,hashisho2019}, and transfer-domain based on the Generative Adversarial Networks (GANs) \citep{islam2019}. Recent approaches based on autoencoders use synthetically paired datasets, which was built and made available in \citep{li2021uda,yufei2020}. Besides, these datasets contain paired images from different water types, as well as additional information, such as: depth scene maps and water attenuation factors. Thereby, specific procedures and prior image processing are required in order to obtain this additional information. \par

In these restoration approaches, the quality of the resulting reference images depends on the ability of the synthesizing method. A good enough approximation of these synthesized images to the real non-degraded one's impacts on the resulting improved images. The dehazed reference image cannot be obtained in complex UW environments, unless standard color boards are taken into the UW scene \citep{wang2019review}. The true intensity and hue of the colors of the objects in the UW scene are unknown in most cases. In the absence of real ground-truth, the quality of the results is defined mainly from the subjective analysis of visual perception. The above considerations show that UW image restoration is a challenging task, and an ill-posed problem. \par

On the other hand, a degradation action on the image can be more easily performed. The UW images tend to present some level of degradation that depends on the environmental conditions in which it were acquired. The degradation itself consists in available information, and it can be estimated and used to increase the degradation of the image \citep{hashisho2019,cli2020}. Unlike restoration, which implies an approximation of the real scene, the degradation task starts from a known condition, and the degraded image can even be evaluated in comparison to the original one.

Deep learning methods for UW image improvement are consistently represented by GAN-based proposals, particularly those involving the style-transfer and transfer-domain approaches. The assumption that some UW images present no degradation level is assumed in the transfer-domain methodologies \citep{fabbri2018,hashisho2019,islam2019}. However, the selection of the images is based on subjective analysis, focused on scenes with objects very close to the camera, and limited or absent background scenarios. This approach can lead to a reduced generalization ability of the method. Style-transfer approaches use outdoor or indoor images in order to train neural networks. Meanwhile, the chosen outdoor/indoor images should bear some feature-similarity to the UW images. This issue improves generalization ability and stability when training neural networks. Despite these issues, the methods also show complex network structures with moderate to high computational cost, requiring synthetic paired datasets, and elaborated training procedures \citep{hashisho2019} \citep{li2019}. \par 

In this study, we propose a self-supervised learning strategy for UW image enhancement. The method is based on a single-image architecture using an autoencoder. The proposed algorithm does not require a ground-truth, and the basic conception comprises the teaching of a network to enhance images from a harsher penalty. This harsher penalty results from the replacement of the output image in the loss function with a synthetically more degraded version. The main assumption considers that a degradation method that captures the degradation nature, and intensifies it, tends to generate more realistic images than a restoration method. In addition, we present an attention mechanism to reduce saturated regions in the enhanced image. These regions surge in images with a strong unbalance of the color channel, and regions with high-intensity pixels related to the average value of the image. \par

The main contributions of this paper are:

\begin{enumerate}
   \item We propose an effective enhancement method for UW images based on a small-sized neural network, a small dataset, and a quite simple algorithm;
   \item We present an alternative approach to enhance natural UW images based on degradation content of the image using a self-supervised learning strategy. The algorithm uses the degradation estimated from the input image to \textit{mislead} the training of an autoencoder to compensate for the degradation, without pre-processing of the image or synthesized ground-truths;
   \item We propose an attention mechanism oriented to limit high intensity regions generated in the enhanced images when the original one presents strong unbalance in color channels and outlier pixels.
\end{enumerate}

In the next sections, we will describe our method and results. In Section 2, the related works are discussed. Section 3 presents the framework for image enhancement and the degradation algorithm. In Section 4, the experimental results are presented, and Section 5 is dedicated to the conclusion.

\section{Related works}

In recent years, the enhancement of the UW images has experimented strong evolution \citep{fayaz2020,wang2019review}. Most of the presented works is related to deep learning methodologies. The main issues tackled by these works are color shift and contrast reduction on images.

The IFM-based enhancement method for UW images presented in \citep{song2020} uses statistical models to estimate the background light from a manually annotated database, and histogram distributions of the UW images. The transmission and scene depth maps were obtained via a variant of the UW Dark Channel Prior (UDCP) \citep{drews2016}, and it was applied to modify the color channel transmission map.  

A method for adaptation of color and contrast enhancement based on digital image processing was presented in \citep{li2022acce}. The algorithm used Gaussian and bilateral filtering to decompose the image into high and low frequency components. The minimization of the difference between the guided image obtained from the low frequency component, and the output image. The enhanced image is combined with the soft-thresholded high frequency component to compose the output image. The method presented effectiveness in the enhancement task. However, it showed significant sensitivity to the parameters for the minimization procedure. The previous selection of the parameters is performed empirically.

IFM-free enhancement methods do not require specialized UW conditions or scene depth information. In \citep{ancuti2012}, an enhancement method based on the fusion technique was presented. The method fuses color corrected and contrast-enhanced versions of the input image to obtain the output image. Despite the effective results, some images show over-enhancement and unnatural colors. In \citep{fabbri2018}, an improvement of the model shown in \citep{li2018a} was presented, where the Underwater GAN (UGAN) was trained to generate turbid images from clear ones. A synthetic dataset is produced, and used to train another GAN. During inference, the generator network predicts clear images from blurry images as input. These methods present unrealistic color correction due to the lack of true colors in the datasets for specific UW scenarios. \par

The study presented in \citep{hashisho2019} introduced the Underwater Denoising Autoencoder (UDAE) model based on the U-Net architecture \citep{ronneberger2015}. Underwater images are presented to the CycleGAN generative model \citep{zhu2017}, for style-transfer between clean and distorted UW images, and to generate a paired UW image dataset. UDAE is trained to restore the image colors. This method depends on the diversity in nature and intensity of the degradation, present in the selected images. 

In \citep{islam2019}, a conditional GAN-based model (FunieGAN) conceived for real-time UW image enhancement was presented. The transfer-domain method developed in \citep{fabbri2018} was used to generate synthetic paired dataset. Then, this dataset was used to train the model. The authors reported a loss of effectiveness for enhancing texture-less and low-contrast images. Also, it was reported that the model is prone to instability during training. 

In a different way, a hybrid approach for UW image restoration with edge-enhancement was presented in \citep{pan2019}. This method uses a CNN to estimate the IFM transmission map of UW images. In addition, a white balance strategy was developed to remove the color cast, and the non-subsampled Contourlet Transform was used to perform denoise and edge enhancement. The neural network was trained using clear UW images, and degraded versions obtained from IFM. However, this algorithm requires prior estimation of background light and filtering in a post-processing step.  

A model based on UW scene priors was introduced in \citep{cli2020}. The IFM was used to generate synthetic UW images and a paired dataset was built. This resulting dataset contains images of several water types, as well as degradation levels, and it was used to train a lightweight neural network to enhance UW images and videos. However, the resulting images required post-processing in order to restore the dynamic range and colors of the images. Furthermore, in \citep{yufei2020}, an approach which combines CNN and IFM was developed. The method divides the image restoration process into two stages, the horizontal one, which embeds the IFM in the neural network, and the vertical one, which restores the image distortion caused by vertical absorption of light. This method requires prior processing of the images in order to estimate the physical parameters, and the image with vertical distortion.

In \citep{li2021uda}, it was presented a model that fuses different feature maps during image representation learning. A synergistic pooling mechanism was used to extract channel-wise attention maps to derive the locally weighted features. The model focused on features related to degraded patches in the UW image in order to improve these patches. Synthetic paired dataset were used to train neural network. The loss function presented a feature loss component, which required previous training of the neural network for feature mapping. In \citep{mello2021} was shown an unsupervised learning methodology based on an encoder-decoder architecture. The neural network presents a double encoder section, and the method estimates the degradation of the image to drive the training. The resulting images show a strong enhancement of the ambient light. Color restoration and dehazing action are limited. Finally, an encoder-decoder neural network integrating different color spaces was developed in \citep{cli2021}. The algorithm uses attention methodology to aim the features' extraction from the images in the RGB, HSV, and Lab color spaces. A transmission map was estimated and used to drive network learning to enhance UW images. The method showed effectiveness, and good performance in the enhancement task. The paired UIEB dataset \citep{cli2020} was used in the training of the neural network. Difficulties in enhancing images in limited lighting were reported by the authors. \par

\section{Methodology}

The perception of the intensities and hues of the real colors is unknown at higher depths, and under limited illumination conditions. In addition, water turbidity generates a loss of contrast and blurring. Methodologies for image improvement based on CNN and CNN-IFM use synthetic paired datasets to achieve or to complement network training.\par

We propose a single-image architecture for UW image enhancement. Our proposal explores the degradation content in the image to drive the enhancement task. The concept relies on the assumption that the UW image presents some level of degradation that can be estimated, and used to drive its reduction or removal from the image. This synthetic degradation action will intensify the degradation level present in the original image. Then, this degraded image will be adopted as the output image of the neural network in the loss function during the training stage. This "trick" will cause a higher penalty in the loss function of a neural network during the training step. Therefore, the neural network will learn to remove the additional degradation by allowing the enhancement of the original image.\par

The proposed methodology uses a fully convolutional autoencoder to perform image enhancement from the described strategy that requires neither synthetic ground-truth nor priors of the image. The following subsection describes the conceptual architecture and the proposed algorithm.

\subsection{Proposed framework for image enhancement}

\begin{figure}[!h]
    \centering
    \includegraphics[width=8.5cm, height=4.8cm]{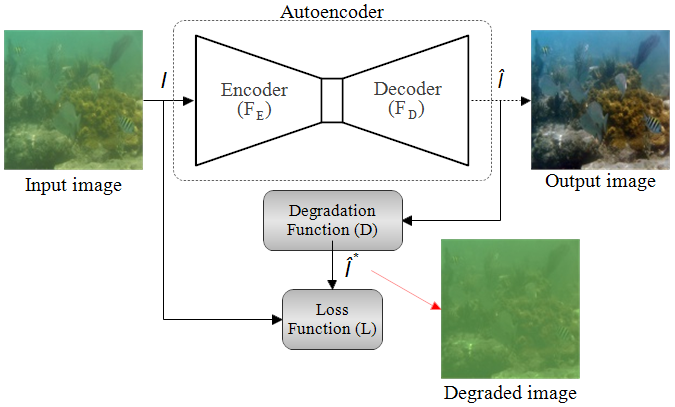}
    \centering
    \caption{Diagram of the conceptual structure of the proposed method.}
    \label{fig:concept}
\end{figure}

Essentially, an autoencoder learns to reconstruct its input information, and to present an identical version of the output. We propose to insert a degradation function $D$ over the training stage that increases the distortion present in the output image of the autoencoder. The conceptual structure of the model is shown in Fig. \ref{fig:concept}.

The resulting image $\hat{I}^*$ from $D$ is a degraded version of the output image $\hat{I}$. The degraded image $\hat{I}^*$ replaces the output image in the loss function. The neural network interprets that its output is not good enough, and learns to correct it. However, effective correction occurs if the degradation is feature-coherent in nature and intensity with those present in the input image. \par
The proposed methodology requires a lower level of supervision compared to peering approaches, since the input image acts as a reference image, not requiring a paired dataset.

\subsection{Degradation Function}

The distortion must be coherent in nature and intensity with the original image. According to Fig. \ref{fig:concept}, the degraded image $\hat{I}^*$ can be described by

\begin{equation} \label{dis1}
    {\hat{I}^*(x,y)} = {\hat{I}(x,y) + \Delta(x,y)},
\end{equation}

\noindent where $\hat{I}$ represents the output image of the autoencoder, and $\Delta$ represents the imposed degradation over $\hat{I}$ by the degradation function $D$. Assuming that a non-distorted image $I_c$ can be defined by removing the distortion $\Delta$ from a degraded image $I^*$; then, it follows

\begin{equation} \label{dis2}
    {I_c}(x,y) = {{I^*}(x,y) - \Delta(x,y)},
\end{equation}

\noindent isolating $\Delta$ in (\ref{dis2}), and using in the (\ref{dis1}), generalizing with the representation of the color channel $(k)$, we have:

\begin{equation} \label{deg_eq}
    {\hat{I}_{(k)}^*}(x,y) = {\hat{I}_{(k)}(x,y) + I_{(k)}^*(x,y) - {I_{c(k)}(x,y)}},
\end{equation}

\noindent where $k \in \{R, G, B\}$. \par

The expression in (\ref{deg_eq}) summarizes the degradation function. The image $\hat{I}_{(k)}^*$ corresponds to the degraded image resulting from the degradation function $D$ in Fig. \ref{fig:concept}. The $I_{(k)}^*$ and $I_{c(k)}$ images are degraded and stretched histogram versions of the input image, respectively. Essentially, the difference between these images in (\ref{deg_eq}) defines the degradation action. It implies that degradation added to the $\hat{I}_{(k)}$ is obtained from two opposite interpretations. First, the image $I_{(k)}^*$ must provide additional degradation-related features that will be added to $\hat{I}_{(k)}^*$.

Second, $I_{c(k)}$ must contain features that improve image quality, and show a path of decreasing degradation. At the same time, the degradation procedure increases the low quality-related features, and decreases the high quality-related features of the image. In the next sections will be presented the procedures to calculate these images. The calculations of these images use an image description inspired by the IFM, and are described in RGB space color.

\begin{figure*}[t]
    \centering
    \includegraphics[width=18cm, height=7.3cm]{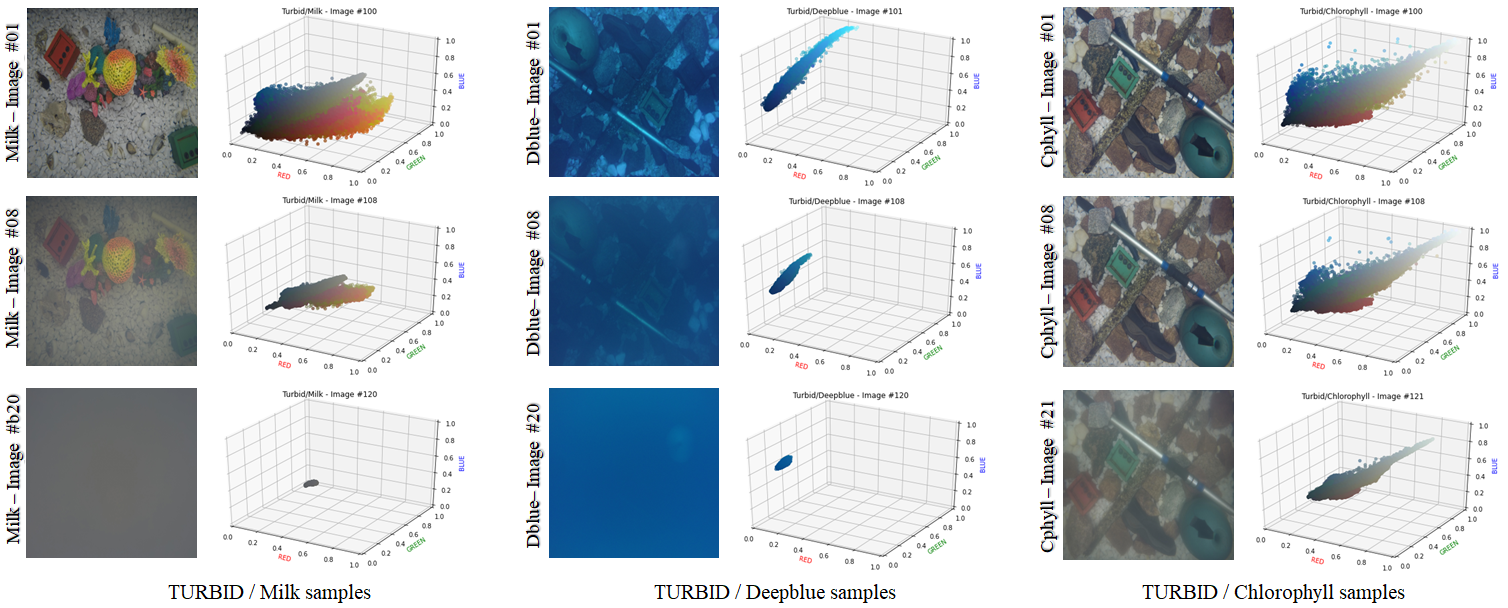}
    \centering
    \caption{Samples of the images from TURBID dataset, and respective pixel distribution in the RGB color space.}
    \label{fig:Turbid_cs}
\end{figure*}

\subsection{Description of the image in the RGB color space}

The image represented by the IFM has embedded elements that describe the image with its related features. An algorithm that uses this model needs to estimate the scene reflectance, water attenuation coefficient, and background light for the image description \cite{berman2019,akkaynak2018}. \par

The proposed algorithm imposes distortion in a coherent manner avoiding over or under degradation. However, the method does not perform any prior processing of the images, and the unique information available is the input image. We adopted an image description contextualized in the color space, but inspired by the physical model presented in the IFM. This description is presented in (\ref{modelo}), and it is used in the estimation of the $I_{(k)}^*$, as shown in (\ref{deg_eq}). 

\begin{equation} \label{modelo}
    {I_{(k)}(x,y)} = {I_{J(k)}(x,y)e^{-gd_{(k)}(x,y)} + (1 - e^{-gb_{(k)}(x,y)}).{\Lambda}_{(k)}},
\end{equation}

\noindent where $I_{(k)}$ is an UW image, $I_{J(k)}$ represents the unknown UW scene without distortion, and $\Lambda_{(k)}$ is defined as the context luminosity of the image. The $gd_{(k)}$ and $gb_{(k)}$ parameters represent turbidity-distance factors related to the water attenuation of the light from the objects in the scene and the ambient light, respectively. These parameters are obtained from the input image in a pixel-wise context. In order to calculate them, we analyzed the characteristics of the UW images in color space under variable turbidity. To this end, we performed a study on the images from the TURBID Dataset \citep{duarte2016}. This dataset contains three sets of real UW images with progressive turbidity: Milk, Deepblue, and Chlorophyll. The Milk subset has 20 images; the Deepblue and Chlorophyll subsets have 21 images. \par

In order to facilitate the analysis of the images, we define the \textit{dynamic range} $b_{w(k)}$ of the image as the difference between the maximum and minimum values by color channel as defined in (\ref{bw}). The maximum value is represented by $max(\cdot)$, and the minimum value by $min(\cdot)$. 

\begin{equation} \label{bw}
    {b_{w(k)}} = {max(I_{(k)}(x,y)) - min(I_{(k)}(x,y))}.
\end{equation}

\subsection{Effects of the turbidity in UW images}

Samples of the images from Milk, Deepblue, and Chlorophyll datasets are shown in Fig. \ref{fig:Turbid_cs}. Also, it presents the distribution of the pixels in the RGB color space for each image. The first perception is image surface shrinkage in the color space with increasing turbidity. The \textit{dynamic range} of the image decreases when the turbidity increases. Specifically, the maximum value is reduced due to forward scattering, and absorption phenomena generated by turbidity. In contrast, the minimum value increases due to the interaction between ambient light and turbidity. This effect increases the perception of turbidity in the image captured by the camera. In Fig. \ref{fig:Turbid_cs}, the \textit{dynamic range} decreases, and the scene converges to the background light, then turbidity perception dominates the image. This is perceptible in all sets of images but most evident in the Milk samples.

\begin{figure}[!h]
    \centering
    \includegraphics[width=\columnwidth]{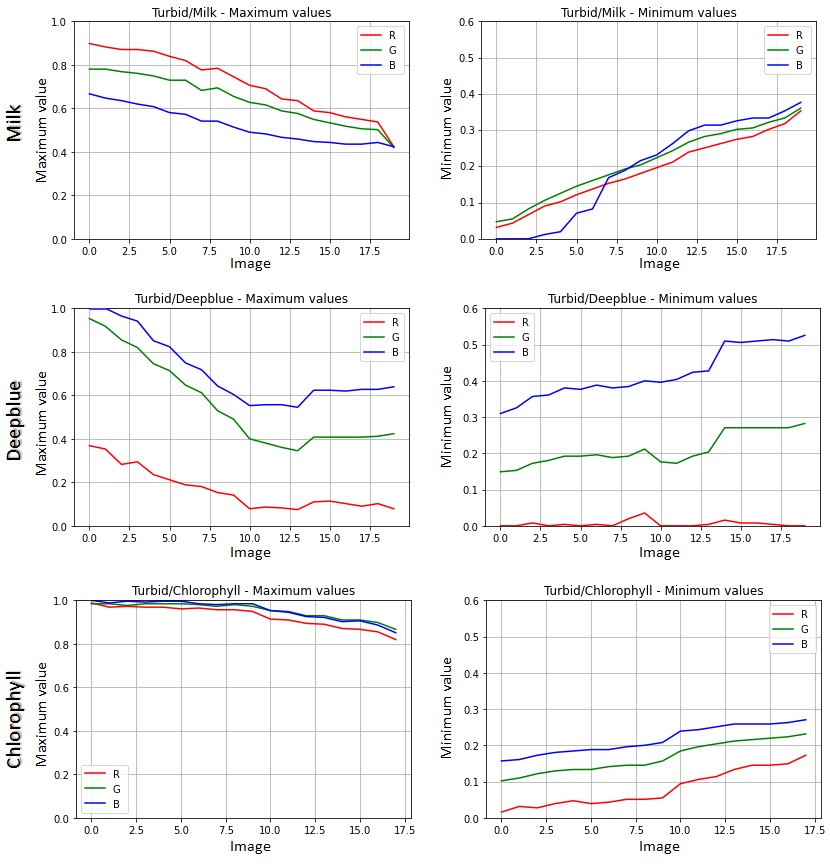}
    \centering
    \caption{Variation of the minimum and maximum values with turbidity for TURBID dataset.}
    \label{fig:maxmin}
\end{figure}

The behavior of the minimum and maximum values is shown in Fig. \ref{fig:maxmin} for the TURBID dataset. The horizontal axis indicates images in ascending order of turbidity. \par

The Milk and Chlorophyll datasets show a balanced color cast and brown-greyish coloring. The effect on the maximum and minimum values presents a uniform variation related to turbidity. The Deepblue dataset presents a strong bluish color cast that accelerates the attenuation of the maximum values, mainly the red and green channels. The backscattering phenomenon is the main cause of the turbidity perception in UW images \citep{akkaynak2018}, and this effect is increased by the color cast.

\subsection{Parameters for description of the image in the color space}

The distorted image generated by degradation function, shown in (\ref{deg_eq}), must have a narrower \textit{dynamic range}. In other words, lower maximum and higher minimum values related to the input image (UW image). This conception is in accordance with the discussion above. \par

The degradation to be added to the output image of the autoencoder is described by the estimation of the $gd_{(k)}$, $gb_{(k)}$, and $\Lambda_{(k)}$ in (\ref{modelo}). The perception of turbidity in the UW image is proportional to the pixel intensity. Thus, the degradation imposed on the pixel must be proportional to the pixel intensity. We used the interaction between the unused interval available to the \textit{dynamic range} $[0,1]$, and the information content inside the \textit{dynamic range} ($b_{w(k)}$). Higher turbidity implies lower $b_{w(k)}$, and less information available in the image. Thus, the parameter $gd_{(k)}$ defines the turbidity effects over the reflectance of the objects in the scene. It is defined as indicated in (\ref{gdi}). The term $[max(I_{(k)}) - I_{(k)}].[I_{(k)} - min(I_{(k)})]$ concentrates the degradation action within the \textit{dynamic range}. It prevents the formation of saturated or zero-intensity regions in the degraded image, and loss of information. This parameter is described in a pixel-wise context, but the pixel coordinates (x,y) are omitted in (\ref{gdi}) for clarity concerns.

\begin{equation} \label{gdi}
    {gd_{(k)}} = {\frac{1 - b_{w(k)}}{b_{w(k)}}[max(I_{(k)}) - I_{(k)}].[I_{(k)} - min(I_{(k)})]}.
\end{equation}

Similarly, the parameter $gb_{(k)}$ is calculated from the interaction between the unused interval available to the \textit{dynamic range} $[0,1]$, the interval $b_{w(k)}$ and the Context Luminosity information $[max(I_{(k)}) - I_{(k)}].[\Lambda_{(k)} - min(I_{(k)})]$ inside of the \textit{dynamic range}. This concept is inspired by the revised image formation model, presented in \citep{akkaynak2018}, where the attenuation factor of the water has different dependencies in the scene and ambient light components in the revised physical model. Thus, $gb_{(k)}$ defines the backscattering effect due to ambient light, and its calculation is expressed in (\ref{gdb}). The pixel coordinates (x,y) are also omitted for clarity.

\begin{equation} \label{gdb}
    {gb_{(k)}} = {\frac{1 - b_{w(k)}}{b_{w(k)}}[max(I_{(k)}) - I_{(k)}].[\Lambda_{(k)} - min(I_{(k)})]}.
\end{equation}

The Context Luminosity ($\Lambda_{(k)}$) is estimated based on the background light definition presented in \citep{akkaynak2018}. Considering the turbidity dominating the scene of an image, in this case, $gd_{(k)}\rightarrow\infty$, and there is no perception of the objects. This event corresponds to $b_w{(k)}\rightarrow 0$, and both color channel median and average values of the image converge to the center of the \textit{dynamic range}. From the Gray World Theory \citep{ebner2007}, the average color in an image with a uniform color distribution can be calculated by the average intensity of the pixels. However, UW images can present irregular light distribution or \textit{light spots} \citep{jian2021}. These phenomena are caused by turbidity, sunlight channeling, and artificial light sources that produce regions with high intensity pixels that bias the averaging. These \textit{outlier} regions affect model performance resulting in images with incorrect ambient light and irregular color enhancement. Thus, the median per channel was adopted to calculate Context Luminosity. This option is due to the median value presenting more robustness to the outlier values than the average value. The Context Luminosity calculation is

\begin{equation} \label{Boo}
    {\Lambda_{(k)}} = median(I_{(k)}(x,y)).
\end{equation}

\subsection{Images for the degradation function}

The degradation algorithm is summarized by (\ref{deg_eq}), and the distortion imposed on the output image of the autoencoder $\hat{I}$ is obtained from the difference between $I_{(k)}^*$ and $I_{c(k)}$. These images comprise different features of distortion, and should produce a reduction in color and contrast along with increased turbidity and colorcast. \par

The $I_{(k)}^*$ image contains the turbidity and colorcast features added to the $\hat{I}_{(k)}$, and it is estimated in two steps using (\ref{modelo}), as follows,

\begin{equation} \label{ideg}
    {I_{(k)}^*}(x,y) = {I_{dbn(k)}(x,y).e^{-gd_{(k)}^*(x,y)} + (1 - e^{-gb_{(k)}^*(x,y)}){\Lambda}_{(k)}^*},
\end{equation}
\noindent where
\begin{equation} \label{idb}
    {I_{dbn(k)}}(x,y) = {I_{(k)}(x,y).e^{-gd_{(k)}(x,y)} + (1 - e^{-gb_{(k)}(x,y)}){\Lambda}_{(k)}}.
\end{equation}

%I_{(k)}.b_{w(k)} - min_{(k)}(I_{(k)})
The image $I_{dbn(k)}$ is a degraded version of the input image $I_{(k)}$ and the parameters $gd_{(k)}$, $gb_{(k)}$ and $\Lambda_{(k)}$ are obtained using (\ref{gdi}), (\ref{gdb}) and (\ref{Boo}), respectivaly. This image defines a new and increased context of degradation, allowing the calculation of the parameters $gd_{(k)}^*$, $gb_{(k)}^*$, and $\Lambda_{(k)}^*$. These parameters are used to calculate $I_{(k)}^*$ in (\ref{ideg}), and they are also calculated by (\ref{gdi}), (\ref{gdb}), and (\ref{Boo}), respectively, but using values from $I_{dbn(k)}$ context.\par

\begin{figure}[h]
    \centering
    \includegraphics[width=5.8cm, height=6.2cm]{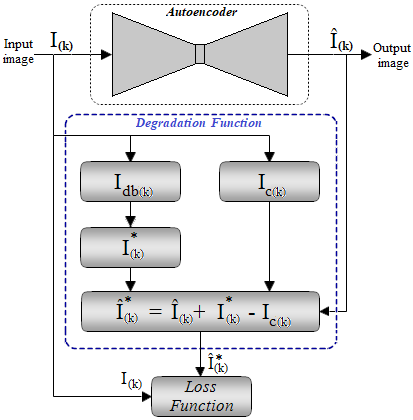}
    \centering
    \caption{Diagram of the method indicating the degradation function and the autoencoder.}
    \label{fig:df}
\end{figure}

The image $I_{c(k)}$ is a histogram-stretched version of $I_{(k)}$, and it is defined in (\ref{inorm}). Essentially, this histogram stretching produces increased contrast and pixel intensity by \textit{dynamic range} expansion \citep{burger_book}. This operation is used in contrast adjustment in non-underwater images, and produces improvement in visual perception of the image. However, UW images tend to present reduced \textit{dynamic range}, and the \textit{stretching} can produce higher gaps between adjacent frequency sets in the histogram. This condition drives higher or excessive texturing effects, and loss of quality in visual perception. Moreover, images with outliers regions and strong unbalance of the color channels tend to present saturated regions when the histogram stretching operation is performed in order to calculate the $I_{c(k)}$. In our algorithm, this image is used as a distortion element for contrast and color, since it is subtracted from $\hat{I}_{(k)}$. Fig. \ref{fig:df} shows the block diagram of the degradation function and the autoencoder.

\begin{equation} \label{inorm}
   {I_{c(k)}(x,y)} = \frac{I_{(k)}(x,y) - min(I_{(k)}(x,y))}{b_{w(k)}},
\end{equation}

\begin{figure}[h]
    \centering
    \includegraphics[width=6cm, height=3.5cm]{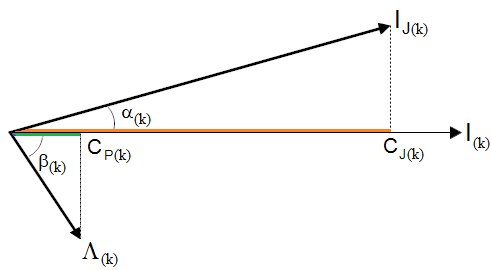}
    \centering
    \caption{Image representation described in (\ref{modelo}) as a vector sum. The image $I_k$ is obtained from the composition between vectors $I_{Jk}$ and $\Lambda_k$. The $C_{Jk}$ and $C_{Pk}$ are the respective components in the $I_k$ direction.}
    \label{fig:vetores}
\end{figure}

\subsection{Loss function}

The loss function adopted is composed of two terms. A term related to the scene radiance context, and another related to the entire image radiance. The scene component is dedicated to highlighting the objects in the scene, and the image component provides global information during the image reconstruction process by the autoencoder.

The scene radiance component is defined from the interpretation of (\ref{modelo}) as a vector sum. Thus, the terms $I_{J(k)}(x,y)e^{-gd_{(k)}(x,y)}$ and $(1 - e^{-gb_{(k)}(x,y)})\Lambda_{(k)}$ compose a complementary action weighted by the exponential factor. We define the image as a vector resulting from these two terms. Assuming that $I_{(k)}$ is the outcome of the vector composition of the $I_{J(k)}$ and $\Lambda_{(k)}$, as shown in Fig. \ref{fig:vetores} and described as

\begin{equation} \label{cossenos}
    {I_{(k)}} = {I_{J(k)}cos(\alpha_{(k)}(x,y)) + \Lambda_{(k)}cos(\beta_{(k)}(x,y))},
\end{equation}

\noindent where:

\begin{equation} \label{alfa}
    {cos(\alpha_{(k)}(x,y))} = {e^{-gd_{(k)}(x,y)}},
\end{equation}
and
\begin{equation} \label{beta}
    {cos(\beta_{(k)}(x,y))} = {(1 - e^{-gb_{(k)}(x,y)})}.
\end{equation}

In addition, all the quantities are defined in the RGB channels. 

The components $C_{Jk}$ and $C_{Pk}$ of the $I_k$ can be expressed by

\begin{equation} \label{CJ}
    {C_{J(k)}(x,y)}  =  {I_{J(k)}(x,y).cos(\alpha_{(k)}(x,y))},
\end{equation}
\begin{equation} \label{CP}
    {C_{P(k)}(x,y)} = {\Lambda_{(k)}.cos(\beta_{(k)}(x,y))},
\end{equation}
and
\begin{equation}
    {I_{(k)}(x,y)}  =  {C_{J(k)}(x,y) + C_{P(k)}(x,y)}.
\end{equation}

Expanding to the degraded image ${\hat{I}^*}$, it follows
\begin{equation}
    {\hat{I}_{(k)}^*(x,y)}  =  {{\hat{C}_{J(k)}}^*(x,y) + {{\hat{C}_{P(k)}}^*(x,y)}},
\end{equation}

\noindent the $C_{P(k)}$ and ${\hat{C}_{P(k)}}^*$ components are related to the environment light that reaches the camera. The $C_{J(k)}$ and ${\hat{C}_{J(k)}}^*$ components prioritize the information about objects in the scene. Essentially, the images $I_{(k)}$ and $\hat{I}_{(k)}^*$ are matched inside the loss function. Instead of using only the $I_{(k)}$ and $\hat{I}_{(k)}^*$ in the loss function, we also use a term composed by $C_{J(k)}$ and ${\hat{C}_{J(k)}}^*$ components as described by (\ref{cc1}) and (\ref{cc2}), respectively.

\begin{equation} \label{cc1}
    {C_{J(k)}(x,y)} = {I_{(k)}(x,y) - C_{P(k)}(x,y)},
\end{equation}
and 
\begin{equation} \label{cc2}
    {{\hat{C}_{J(k)}}^*(x,y)} = {\hat{I}_{(k)}^*(x,y) - {\hat{C}_{P(k)}}^*(x,y)}.
\end{equation}

The $C_{P(k)}$ and ${\hat{C}_{P(k)}}^*$ components are calculated using (\ref{CP}) with the respective parameters. The term of the loss function is obtained from the Mean Square Error (MSE) between the components, which is calculated as

\begin{equation} \label{loss0}
   {\mathcal{L}_{sc}} = MSE{(C_{J(k)}(x,y), \hat{C}_{J(k)}^*(x,y))}.
\end{equation}

The term of the image radiance is composed by the MSE between images $I_{(k)}$ and $\hat{I_{(k)}}^*$ as indicated in (\ref{loss00}). This term provides image context information avoiding the formation of artifacts in the output image due to loss unbalance, which appears when employing only scene components.

\begin{equation} \label{loss00}
   {\mathcal{L}_{im}} = MSE{(I_{(k)}(x,y), {\hat{I}_{(k)}^*(x,y))}}.
\end{equation}

The scene and image radiance terms described in (\ref{loss0}) and (\ref{loss00}) are used to compose the loss function used in training the model, as indicated in (\ref{loss}). The constants $c_1$ and $c_2$ are used to weight the terms. They are indicated in the Experimental Results Section.

\begin{equation}  \label{loss}
   {\mathcal{L}} = {c_1.}{\mathcal{L}_{sc}} + {c_2.}{\mathcal{L}_{im}},
\end{equation}

\noindent where the $c$ factors are weights for loss terms.\par

\subsection{Attention for saturated regions in the image}

\begin{figure}[h]
    \centering
    \includegraphics[width=8.2cm, height=4.cm]{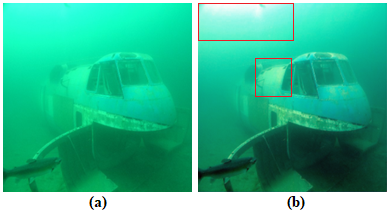}
    \centering
    \caption{Example of image with partial and strong incidence of the light source causing unbalance in color channels and outlier regions. The generated high-intensity areas in the output image are indicated inside the red rectangles. (a) Input image. (b) Output image.}
    \label{fig:outlier1}
\end{figure}

The histogram stretching performed on images presenting strong unbalance of the color channel and outlier regions, simultaneously, tends to generate high-intensity or saturated areas in the images. The unbalance of the color channels can be caused by the nature of the scene itself or by the turbidity and color cast effects. The outlier regions are generated by the presence of pixels with an intensity that is much higher than the average of the image. They are due the irregular ambient light distribution. Another characteristic of these images is the much lower minimum values. Under these conditions, the histogram stretching operation generates high values for the high-intensity pixels in the normalization procedure. Thus, saturated areas are created in the histogram-stretched image. The content of this image is removed from the output image in the degradation function, including the saturated area. During the training phase, the autoencoder learns to compensate for the degradation. In this case, it generates compensation for the removed saturated areas, and these areas are formed in the output and enhanced images. Fig. \ref{fig:outlier1} shows an example of presenting saturated and high-intensity areas. The red rectangle highlights the saturated area generated by the method. In order to overcome this drawback, we propose an attention mechanism that detects these outlier pixels and color channel unbalance. The algorithm was implemented in the degradation function, and compensates for these effects. It is discussed in the following section.

\begin{figure}[h]
    \centering
    \includegraphics[width=5.5cm, height=5.6cm]{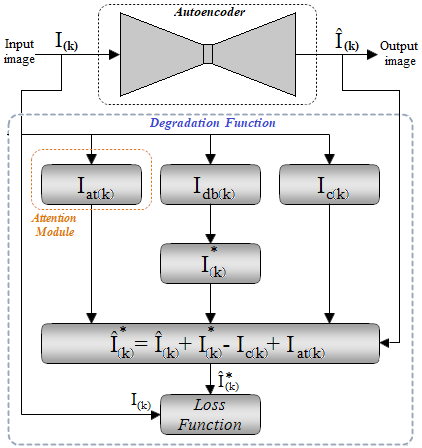}
    \centering
    \caption{Diagram of the method with the attention module inserted in the degradation function.}
    \label{fig:outlier2}
\end{figure}

In order to reduce these saturated regions, we propose an attention mechanism dedicated to treating these regions. The attention module is inserted in the degradation function as an additional term of degradation. The diagram in Fig. \ref{fig:outlier2} shows the inserted attention module. This module maps the input image, and generates the attention image $I_{at(k)}$, containing regions of high intensity generated by the presence of outliers pixels and color channels unbalance. This image $I_{at(k)}$ is added to the output image in the degradation function, acting as additional degradation. The novel degradation function is

\begin{equation} \label{deg_eq_att}
    {\hat{I}_{(k)}^*}(x,y) = {\hat{I}_{(k)}(x,y) + I_{(k)}^*(x,y) - I_{c(k)}(x,y) + I_{at(k)}(x,y)},
\end{equation}

The attention image $I_{at(k)}$ is a version with stretched histogram of the image $I_{a(k)}$, described in (\ref{attention0}). The histogram stretching is performed using (\ref{inorm}) with the parameters of (\ref{attention0}). Thus, $I_{at(k)}$ is

\begin{equation} \label{attention}
    I_{at(k)} = \frac{I_{a(k)}-min(I_{a(k)})}{max(I_{a(k)})-min(I_{a(k)})}.
\end{equation}

The image $I_{a(k)}$ is defined as
\begin{equation} \label{attention0}
    I_{a(k)} = g_{r(k)} \mid av(I_k) - av_g(I_{(k)})\mid \frac{I_{x(k)}}{I_{(k)}},
\end{equation}

\noindent where $g_{r(k)}$ is a reinforcement factor calculated in (\ref{gr}), $av(I_{(k)})$ and $av_g(I_{(k)})$ are the average value per color channel and global average of the image, respectively. The detection of the unbalance of the color channels is performed by the term $\mid av(I_{(k)}) - av_g(I_{(k)})\mid$ in (\ref{attention}). $I_{(k)}$ is the input image, and $I_{x(k)}$ is the image containing outlier pixels, which is calculated as indicated in (\ref{imgx}). The pixels above of threshold represented by $max(I_{(k)})\frac{thr_{(k)}}{av(I_{(k)})}$ are considered outlier pixels. Thus, they are selected to compose $I_{x(k)}$.

\begin{equation} \label{gr}
    g_{r(k)} = {1 + \frac{thr_{(k)}}{av(I_{(k)})}},
\end{equation}

\begin{equation} \label{imgx}
    I_{x(k)} = 
\left\{ 
    \begin{array}{ll}
     I_{(k)} - max(I_{(k)})\frac{thr_{(k)}}{av(I_{(k)})},  & \, I_{x(k)}\geqslant 0 \\
      0,  & \, I_{x(k)}<0,
    \end{array}
\right.
\end{equation}

\noindent where $max(I_k)$ correspond to the maximum value per color channel of $I_{(k)}$. The term $thr_{(k)}$ in (\ref{gr}) and (\ref{imgx}) is defined in (\ref{thr}), and correspond to the threshold value for outlier detection. The presence of outliers in the image increases the distance between the average and the median values, because the average is strongly affected by data outliers. We explore this effect to detect the presence of outliers in the image $I_{(k)}$. The index $(k)$ represents the color channel and $k \in \{R,G,B\}$.

\begin{equation} \label{thr}
    thr_{(k)} = 
\left\{ 
    \begin{array}{ll}
     av(I_{(k)})-mdn(I_{(k)}),  & \, thr_{(k)}\geqslant 0 \\
      0,  & \, thr_{(k)}<0,
    \end{array}
\right.
\end{equation}

\noindent where $mdn_{(k)}$ is the median value per color channel of the $I_{(k)}$.

\section{Experimental results}

The details of the implementation results and performance analysis of the method are described in this section. Also, we present a comparative study involving the proposed method, and other enhancement methodologies.

\subsection{Autoencoder}

In this work, we consider the enhancement task performed by the neural network as an image reconstruction task, driven by the degradation in a self-supervised approach. The neural network must be able to learn hidden structures in unlabeled data, representing them in latent features. A deep learning model based on autoencoder performs this trying to reconstruct its input \citep{pretorius2018}. The choice for an architecture based on autoencoder stems from this conception and the nature of the method. The objective was to implement a neural network with low computational cost and an easy approach for self-learning. \par

\begin{figure}[htb]
    \centering
    \includegraphics[width=8.4cm, height=4cm]{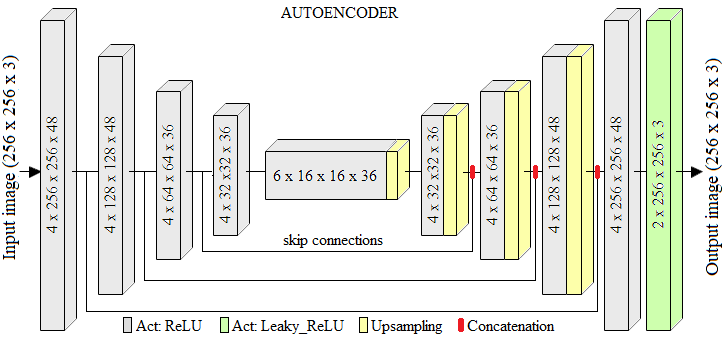}
    \centering
    \caption{Architecture of the implemented autoencoder.}
    \label{fig:AE}
\end{figure}

The implemented autoencoder is shown in Fig. \ref{fig:AE}. The structure of the neural network is a Fully Convolutional Network (FCN) with 170k trainable parameters. The downscaling steps are performed via \textit{strided} convolutions, instead of pooling layers, in order to minimize information loss. The upscale in the decoder section is performed by Keras 2D upsampling layers, with size $(2,2)$, interpolation by nearest neighbors and other arguments with default values. The activation functions are ReLU, except in the final layers of the decoder that use Leaky-ReLU. The convolutional layers use L1 regularization with factors of the kernel, and bias set to $15\times10^{-6}$ and $1.5\times10^{-6}$, respectively. Only the layers of the autoencoder have trainable parameters, and the degradation function was implemented as a non-trainable layer.\par

\subsection{Training details and dataset}\label{training_d}

The autoencoder was implemented in Keras/Tensorflow\textsuperscript{\textregistered}, and it was initialized with Glorot Normal Algorithm. The optimizer is Adam with a learning rate equal to 0.0008, and other parameters were maintained in its default values. In the Leaky-ReLU layers, the alpha parameter is equal to 0.19. The training was performed in 200 epochs with a batch size equal to 6. The computer configuration is i5-6400 CPU, 32 MB RAM with a Titan X GPU. The image format is RGB $256\times256\times3$ scaled to the range $[0, 1]$. \par

As a future prospect for extending the application of the method is in embedded vision systems of Remotely Operated Vehicles (ROV). Processing time requirements for real time operations are an important issue in these systems. Thus, we evaluated the inference time of the trained model for the processing one image of $256\times256\times3$ in 0.25s, in a computer configuration i5-6400 CPU, 32 MB RAM, O.S. Ubuntu 18.04.5, Tensorflow vs.2.20, Keras vs.2.3.1. This results in a rate of 4 FPS, approximately. This value is quite low for real time applications. We expect that with optimization procedures favoring the operating system and model performances, this inference time can be reduced. However, more investigations are needed on this topic.
\par

The dataset used to train the autoencoder contains 2200 real UW images. It was built prioritizing images with medium to high levels of turbidity, reduced color and contrast perception. Also, these present degradation related to shades, and intensities of color cast and ambient light. Images with very low degradation were also used, but not prioritized. We used 800 images from UIEBD database \citep{li2019}, 400 images from the SUIM database \citep{islam2019}, 100 images from the RUIE dataset \citep{liu2020}, and 900 UW images collected from the internet. These mentioned datasets were used in recent studies, related to the UW image enhancement \citep{li2021uda, yufei2020, cli2021, islam2019, cli2020}. The data were randomly split into 90\% for training and validation, and 10\% for test. Cross-fold validation was performed during the evaluation procedures to check for consistency in image reconstruction (autoencoder) with degradation function added to the pipeline. We used an additional dataset containing 90 images for evaluation in the comparative study described below in the respective section. These images were selected from the datasets cited above, and they were  not used in training of the neural network. The $c$ factors values in the loss function expression were empirically defined as $c_1 = 0.65$, and $c_2 = 0.35$.

\subsection{Resulting images from the proposed method}

\begin{figure*}[h!]
    \centering
    \includegraphics[width=18.1cm, height=4.6cm]{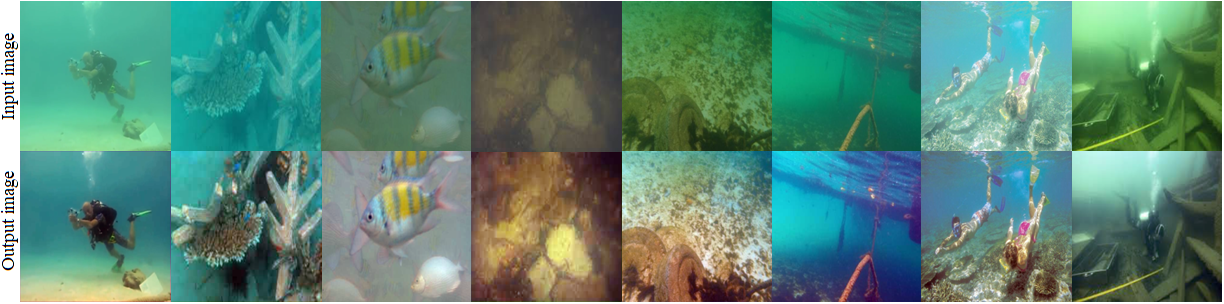}
    \centering
    \caption{Examples of enhanced images obtained from our method.}
    \label{fig:saidas}
\end{figure*}

The resulting images from the proposed method are obtained at the end of the training phase. Samples of the input, output and degraded images are indicated in Fig. \ref{fig:saidas}. The objects in the scene show a natural aspect when compared to the input image. Visual perception points out contrast improvement and color recovery. Also, the images show color preservation furthering the color constancy properties (\citep{ebner2007}). In the context of this work, this color preservation is an important issue, because no synthetic reference images are used. The proposal does not focus on changing the color perception of the image objects. In addition, the method presents good performance in turbidity and colorcast reduction. The reduction in turbidity perception is more evident in foreground regions as a result of the scene-related component in the loss function. \par

Fig. \ref{fig:saidas} shows scenes in distinct ambient light, color cast, and turbidity.
Underwater images bring a myriad of complexities that pose challenging issues, and define the performance of a method under diﬀerent aspects of evaluation. The depth affects color and ambient light in UW images. The type of water is related to the nature and quantity of particles in suspension in the water and impacts on contrast, color cast, and blurring.  Essentially, the method considers these features as a result of the influence of depth and water type on image acquisition. Thus, the available information about ambient light and turbidity is estimated from the input image. These physical quantities are expressed and explored inside the degradation function as color information, driving the enhancement task. The resulting images are shown in Fig.  \ref{fig:saidas} indicate that the methodology is able to enhance images in different UW environments. The model presents a good generalization ability, resulting from the dataset composition, and the nature of the methodology.

\begin{figure*}[ht] 
    \centering
    \includegraphics[width=16cm, height = 14cm]{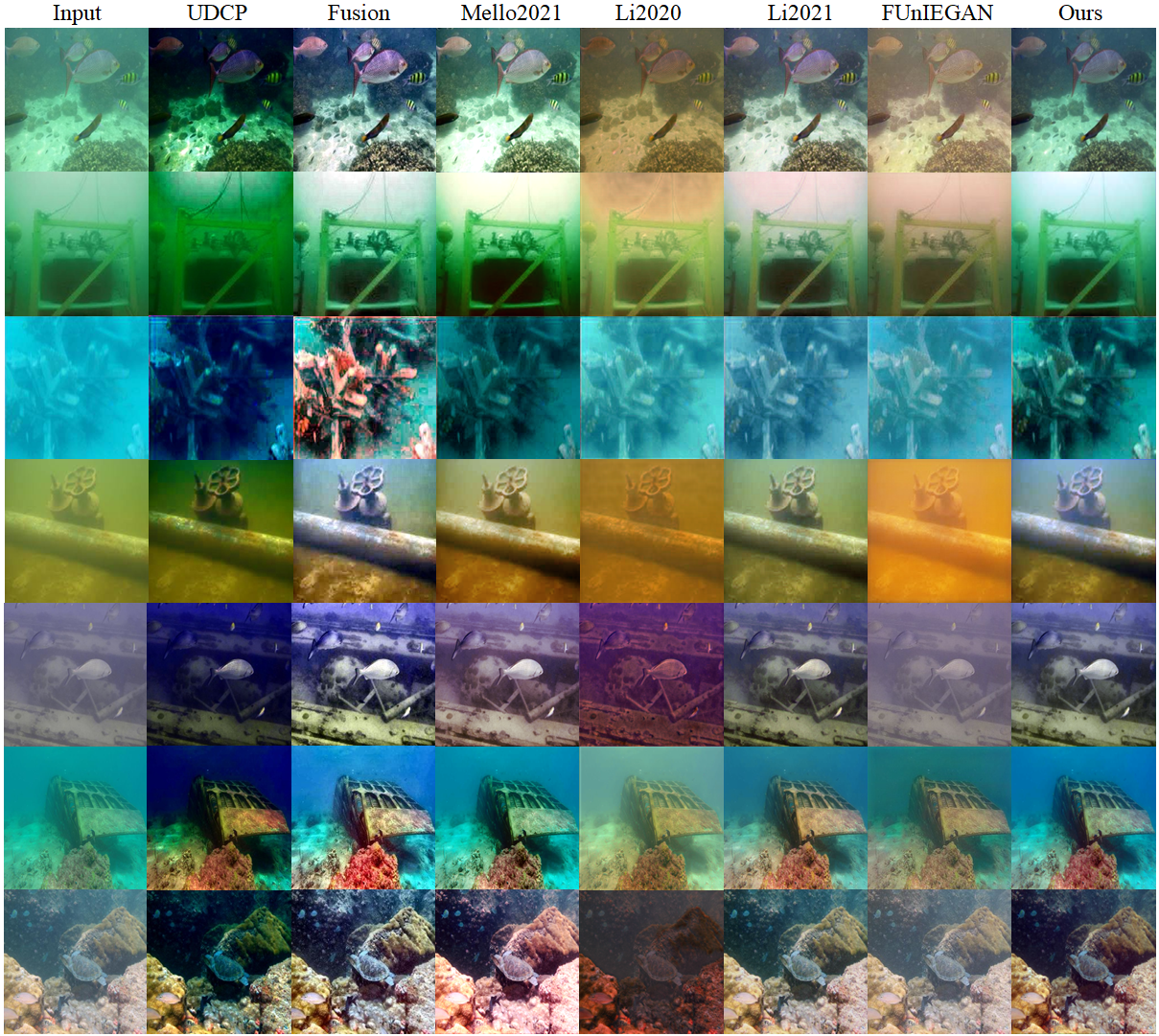}
    \caption{Output images for the comparative study. Methods: UDCP \citep{drews2016}, Fusion \citep{ancuti2012}, Mello2021 \citep{mello2021}, Li2020 \citep{cli2020}, Li2021 \citep{cli2021}, FUnIEGAN \citep{islam2019}, Ours.}
    \label{fig:comparativo}
\end{figure*}

\subsection{Comparative study}

We performed a comparative evaluation of the method related to the consolidated methodologies oriented to UW enhancement and deep learning based methods, developed in \citep{cli2021,cli2020,islam2019, mello2021}. Also, we compared the proposed method to the works presented in \citep{ancuti2012} and \citep{drews2016}, both based on digital image processing. In the comparative analysis, the methods are identified by "Li2021" \citep{cli2021}, "Li2020" \citep{cli2020}, "FUnIEGAN" \citep{islam2019}, "Fusion" \citep{ancuti2012}, "UDCP" \citep{drews2016}, "Mello2021" \citep{mello2021}, and the present proposal by "Ours".

\subsubsection{Qualitative analysis}

For evaluation concerns, we consider as quality criteria the ability of the method in recovery the color and contrast, reduction of the color cast and dehaze action. Also, the color constancy of the resulting images is considered through the color preservation and coherency between the input image (UW image) and the enhanced image. Fig. \ref{fig:comparativo} shows samples of the resulting images obtained from the methods used in the comparative study. These images compose the dataset with 90 samples mentioned in Section \ref{training_d}, and are used in the comparative study only.\par

Most of the methods present an irregular enhancement action with strong color and color cast increasing. The UDCP showed effectiveness in the haze reduction, but with strong color distortion, and darkening in some images. The Fusion method presented an effective dehaze action and contrast recovery, but with strong texturing of the image, and some images showing excessive color enhancement. FUnIEGAN presented low performance dealing with turbidity, and color distortion seemingly associated with the excessive enhancement of the red channel. Li2020 showed intense enhancement of the red channel, and difficulty in leading with haze. Mello2021 presented irregular enhancement action in images with unbalanced ambient light, generating dark or very clear regions in the enhanced images. The method Li2021 and Ours had the best performances in the image improvement. Color preservation and recovery have similar intensities and dehaze action. However, the method Li2021 produced a more natural colorfulness in some images, while Ours showed the best improvement in contrast. Both methods presented good generalization ability.  \par

In our method, the dehazing action is conditioned by the scene components of the loss function. These components drove the enhancement of the scene, and the haze reduction was concentrated in the foreground regions of the image. This scene enhancement provides an important property for UW application-oriented tasks, such as UW maintenance, inspection, and fauna and flora explorations. Turbidity effect reduction over wide, and long-range scenarios can be important to monitoring and inspection activities. In our context, large or restricted scenes are tackled similarly. In acquired images, with scenes showing objects far from the camera, the turbidity perception is more accentuated. The dehazing task performed by the method acts uniformly on the image, but in these regions it is less intense. \par

\subsubsection{Quantitative analysis}

The quantitative analysis of underwater images is a poorly defined issue, due to the lack of consensus on the effectiveness of existing metrics in representing the quality of these images \citep{han2018}. Usually, the authors use an integrated set of the metrics containing the image quality metrics that require or not a reference image (ground-truth). The metrics that use reference or ground-truth images were developed to indoor/outdoor applications \citep{wang2019review}, but applied to UW image analysis, and they need synthesized paired datasets, built from the UW images. On the other hand, there are no-reference metrics proposed specifically for UW image quality analysis. We proceeded with the quantitative evaluation of the images, using the Underwater Image Color Evaluation (UCIQE) \citep{yang2015uciqe},  Underwater Image Quality Metric (UIQM) \citep{panetta2015}, and CCF metric \citep{wang2018CCF}. Also, we use the no-reference metric Natural image quality evaluator (NIQE) \citep{mittal2013}.

The UCIQE measures color degradation in the CIELab color space, based on standard deviation in chroma, average saturation, and difference between extreme values in luminance. The UIQM metric evaluates color degradation with measurements of colorfulness using the opponent color theory, and blurring degradation with measurements of edge sharpness and contrast. UCIQE, UIQM are a linear combination of weights obtained from a subjective test, and related to specific attributes, such as contrast and saturation, besides, luminance and chroma. Higher scores resulting from UCIQE and UIQM indicate higher image quality. Furthermore, CCF measures color degradation with a colorfulness index, blurring, and lack of visibility with a contrast measure and a foggy index. The weighting of these indexes is based on subjective tests, where images of a color chart are captured at different distances and turbidity. Higher values of CCF score indicate better image quality. Moreover, NIQE considers human vision sensitivity to high-contrast areas in images. It uses multivariate gaussian (MVG) to define the feature model of sensitive areas where larger values of the model parameter higher the quality of the image. A smaller score on NIQE indicates better perceptual quality.

Table \ref{table:1} shows the UIQM, UCIQE, CCF, and NIQE scores for the compared methods. The best UCIQE score is UDCP, and the method Li2021 had the best performance in UIQM score. Besides, Fusion presented the best CCF score, while Ours had the best NIQE score.

\begin{table}[ht]
  \centering
\small
  \caption{Quantitative results for UIQM, UCIQE, CCF, and NIQE metrics. UDCP (Drews et al.,2016) \citep{drews2016}, Fusion (Ancuti et al., 2012) \citep{ancuti2012}, Mello2021 (Mello et al., 2021) \citep{mello2021}, Li2020 (Li et al.,2020) \citep{cli2020}, Li2021 (Li et al, 2021) \citep{cli2021}, and FUnIEGAN (Islam, et al., 2019) \citep{islam2019}.}
  %\begin{normalsize}
    \label{table:1}
    \begin{tabular}{ccccc}\hline{\Large \strut}
        Metric & UIQM $\uparrow$ &  UCIQE $\uparrow$ & CCF $\uparrow$ & NIQE $\downarrow$ \\ \hline
        UDCP & 1.604 & \textbf{6.915} & 34.229 & 4.559 \\ %\hline
        Fusion & 2.623 & 5.568 & \textbf{36.895} & 5.875 \\ %\hline
        Mello2021 & 2.189 & 4.934 & 27.803 & 4.786 \\ %\hline
        Li2020 & 2.654 & 3.281 & 13.737 & 4.652 \\ %\hline
        Li2021 & \textbf{2.913} & 3.593 & 19.824 & 4.830 \\ %\hline
        FUnIEGAN & 2.628 & 3.414 & 14.282 & 5.213 \\ %\hline
        Ours & 2.649 & 4.623 & 23.492 & \textbf{4.415} \\ \hline
%        FSIMc $\uparrow$ & 0.717 & 0.813 \\ \hline
    \end{tabular}
  %\end{normalsize}
\end{table}

\begin{figure}[h!]
    \centering
    \includegraphics[width=8.3cm, height=7cm]{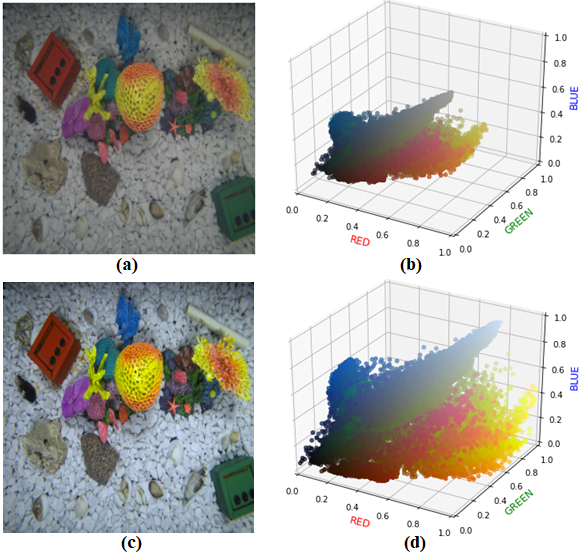}
    \centering
    \caption{The image "ref" of the Milk dataset and the respective stretched-histogram, and white-balanced version used as reference image in the quantitative evaluation with referenced metrics. (a) Original image from the Milk dataset (\#ref), (b) Pixel distribution in the RGB color space of the original image, (c) Expanded dynamic and color balanced version of the image "ref", and (d) Pixel distribution in the RGB color space of the image with expanded dynamic and color balanced version.}
    \label{fig:ref_turbid}
\end{figure}

The values presented in the Table \ref{table:1} point out some issues related to the characteristics of the metrics. The UW image metrics focus on feature intensities like chroma and saturation (UCIQE), and colorfulness and contrast (UIQM), colorfulness, contrast and turbidity (CCF). They tend to produce high scores for images with high contrast and extreme chroma \citep{wang2019review} ignoring human perception context \citep{berman2019}. The results shown in Table \ref{table:1} do not present consistency with subjective evaluation of the images. The UCIQE and CCF scores suggest similar performance in the image assessment. The scores obtained by UDCP are much higher than other ones. They are resulting from excessive color improvement, and darkening perceptible in visual analysis. However, these distortions caused a loss of contrast and color diversity, resulting in the lowest UIQM score. The methods that did not score high on the UCIQE present high values for UIQM. \par

The resulting scores for most methods show uneven performance on the set of metrics. This condition makes a concise evaluation difficult. However, this can be mitigated when we consider the balance of the method scores. In this context, the methods Fusion, Li2021, and Ours presented the best performances. This approach is consistent with the visual perception of the images. These non-reference metrics have a sensibility to different features, and refer to distinct concepts of UW image quality. Underwater image enhancement quality metrics are an open issue, and they lack higher investigations. \par

\subsection{Quantitative analysis using full-reference metrics}

In order to provide a more concise quantitative evaluation of the method, we compared the methods using full-reference metrics often utilized to evaluate image quality. Specifically, Peak Signal-to-Noise Ratio (PSNR), Mean Square Error (MSE) \citep{steffens2020}, Structural Similarity (SSIM) \citep{wang2004}, Gradient Magnitude Similarity Deviation (GMSD) \citep{wufeng2013}, CIEDE2000 \citep{sharma2005}, Feature Similarity, and Feature Similarity with Chrominance (FSIM and FSIMc) \citep{zhang2011}. The evaluation contexts of these metrics are pixel-wise (PSNR, MSE), structure-related (SSIM, GMSD), color (CIEDE2000), and features on visual perception (FSIM, FSIMc). However, all these metrics require a reference image. In this approach, we do not use reference images. Methodologies based on deep learning that use these images, synthesized them using a specific method, and proceed to the training of the their models. We consider that this approach can bias quantitative analysis. Since, a model trained using synthetic and improved images, restored with a specific method, can present good scores in referenced metrics, when using reference images obtained from the same specific method. Thus, we opted to use images from the Milk dataset provided by the Turbid dataset, instead of synthesizing reference images. The Milk dataset has 20 images presenting progressive turbidity, with the first image ("ref") acquired in water without turbidity. We used as reference image a stretched-histogram and white-balanced version of this image, shown in Fig. \ref{fig:ref_turbid}. This version allows improved color and contrast perception, and minimizes the darkened context of the original image, caused by the controlled illumination, used in the image acquisition procedures \citep{duarte2016}. Fig. \ref{fig:ref_turbid} (a) and (c) show the first image from the Milk dataset and the stretched-histogram, and white-balanced version, respectively. The distributions of the pixels in the RGB color space for both images are shown in Fig. \ref{fig:ref_turbid} (b) and (d), respectively.

\begin{figure*}[ht] 
    \centering
    \includegraphics[width=17.2cm, height = 11.2cm]{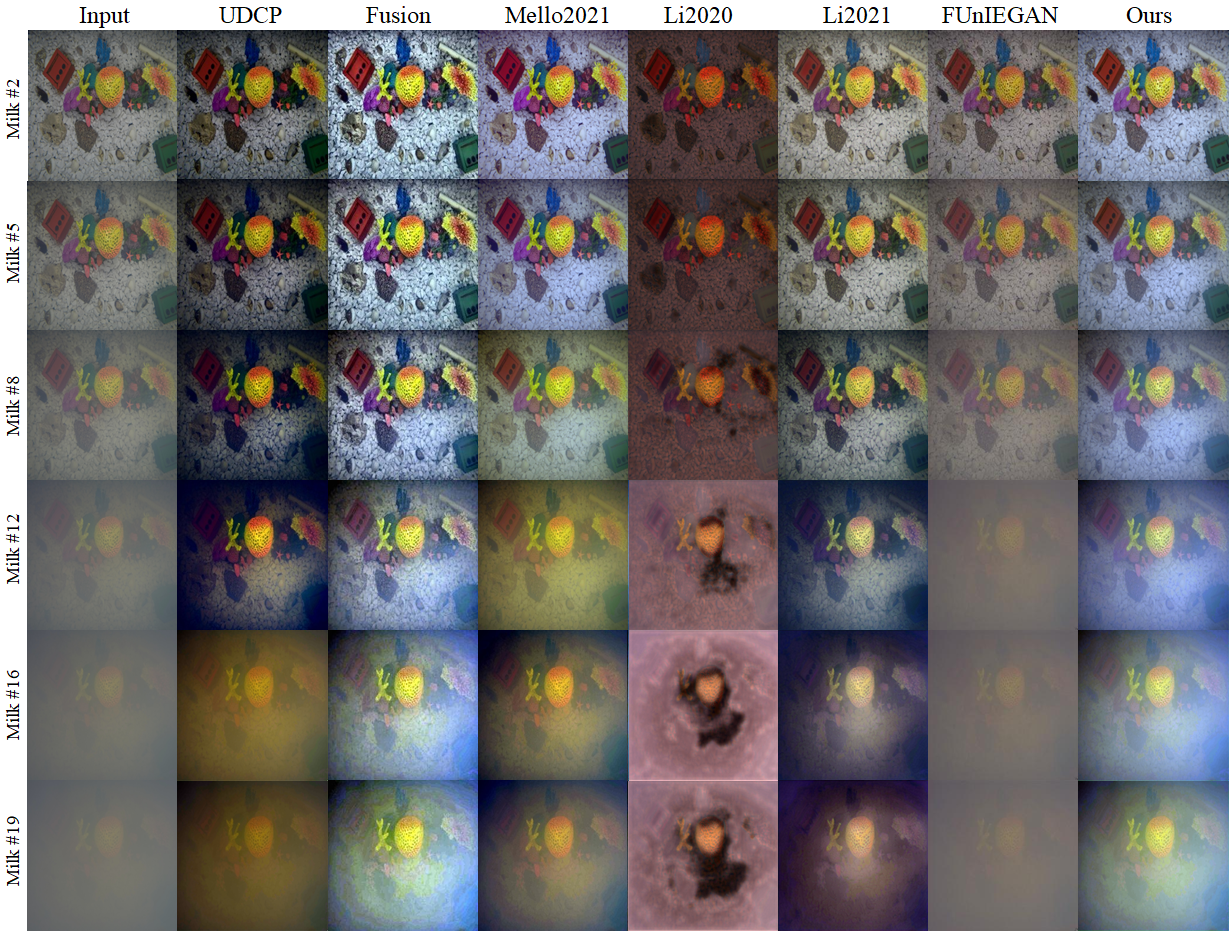}
    \caption{Output images for the comparative study using images from the Turbid/Milk dataset. Methods UDCP \citep{drews2016}, Fusion \citep{ancuti2012}, Mello2021 \citep{mello2021}, Li2020 \citep{cli2020}, Li2021 \citep{cli2021}, FUnIEGAN \citep{islam2019}, Ours.}
    \label{fig:Quant_turbid}
\end{figure*}

Fig. \ref{fig:Quant_turbid} shows samples of the resulting images from the methods for the images of the Turbid/Milk dataset. These images can be used to evaluate methods for progressive turbidity. From the visual perspective, the Fusion and Ours methods performed better under severe turbidity conditions, maintaining the color and scene structure perception. Table \ref{table:2} shows the resulting scores for the image quality metrics that use the reference image. \par

The scores of the metrics related to error and noise (MSE and PSNR) show good performance of Ours, Fusion, and Mello2021. This is consistent with the subjective analysis of the images. These methods showed quite similar results, mainly in images with a low level of turbidity. Our proposal presented the best scores for the scene structure-related metrics (PSNR, MSE, SSIM), and for the feature perception on gray scale (FSIM). These results are important to the visual perception of UW images. The human vision system is less sensible to color changes than luminance changes \citep{burger_book}. The depth and water turbidity effects on the colors are increased under reduced ambient light. Under these conditions, scene perception is driven by luminance. \par

Furthermore, our method also presented the best score in the GMSD metric. Since that the gradient images are sensitive to image distortions, whereas different local structures in a distorted image suffer different degrees of degradation. GMSD captures image local quality, and the standard deviation of the image's gradient map is computed as the final image quality index. \par

The method Fusion presented better performance in color-related metrics (CIEDE2000, FSIMc); but, our method also had high scores in these metrics. Color recovery is an important issue in the enhancement task. Unlike the other deep learning-based methods, our proposal does not use paired datasets; hence, additional information provided by the synthetic ground-truth images is not available. Therefore, our method enhanced the color information present only in the input image. However, we did not consider this condition as a limitation, but a characteristic of the method, since the real intensity and hue of the colors in UW images are unknown.

From the above discussion, our method and Fusion showed the best scores for the referenced metrics. The results of this experiment agreed with the qualitative analysis, based on visual perception of the images, and they are consistent with the qualitative analysis for UW images, discussed in the previous subsection. 

\begin{table*}[ht]
  \centering
%\small
  \caption{Quantitative results for PSNR, MSE, SSIM, CIEDE2000 (CIE2K), FSIM, FSIMc, and GSMD metrics. The adopted reference image is the expanded \textit{dynamic range} and white balanced version of the image 'ref' from Milk dataset, indicated in the Fig. \ref{fig:ref_turbid}(c). UDCP (Drews et al.,2016) \citep{drews2016}, Fusion (Ancuti et al., 2012) \citep{ancuti2012}, Mello2021 (Mello et al., 2021) \citep{mello2021}, Li2020 (Li et al.,2020) \citep{cli2020}, Li2021 (Li et al, 2021) \citep{cli2021}, and FUnIEGAN (Islam, et al., 2019) \citep{islam2019}}
  %\begin{normalsize}
    \label{table:2}
    \begin{tabular}{cccccccccc}\hline{\Large \strut}
        Metric & PSNR $\uparrow$ &  MSE $\downarrow$ & SSIM $\uparrow$ & CIE2K $\downarrow$ & FSIM $\uparrow$ & FSIMc $\uparrow$ & GMSD $\downarrow$ \\ \hline
        UDCP & 10.17 & 0.099 & 0.492 & 29.15 & 0.972 & 0.749 & 34$e^{-6}$ \\ %\hline
        Fusion & 17.06 & 0.024  & 0.704  & \textbf{14.20} & 0.976 & \textbf{0.812} & 26$e^{-6}$ \\ %\hline
        Mello2021 & 16.46  & 0.045  & 0.670  & 20.44  & 0.977 & 0.743 & 24$e^{-6}$ \\ %\hline
        Li2020 & 11.23 & 0.083 & 0.481 & 29.72 & 0.926 & 0.645 & 48$e^{-6}$ \\ %\hline
        Li2021 & 13.95 & 0.054 & 0.650 & 21.63 & 0.975 & 0.788 & 26$e^{-6}$ \\ %\hline
        FUnIEGAN & 14.52 & 0.036 & 0.551 & 18.52 & 0.973 & 0.620 & 37$e^{-6}$ \\ %\hline
        Ours & \textbf{17.65} & \textbf{0.023} & \textbf{0.706} & 15.32 & \textbf{0.979} & 0.763 & \textbf{23$e^{-6}$} \\ \hline
%        FSIMc $\uparrow$ & 0.717 & 0.813 \\ \hline
    \end{tabular}
  %\end{normalsize}
\end{table*}

\subsection{Performance of the method with attention module}

\begin{figure}[h]
    \centering
    \includegraphics[width=8.7cm, height=7.4cm]{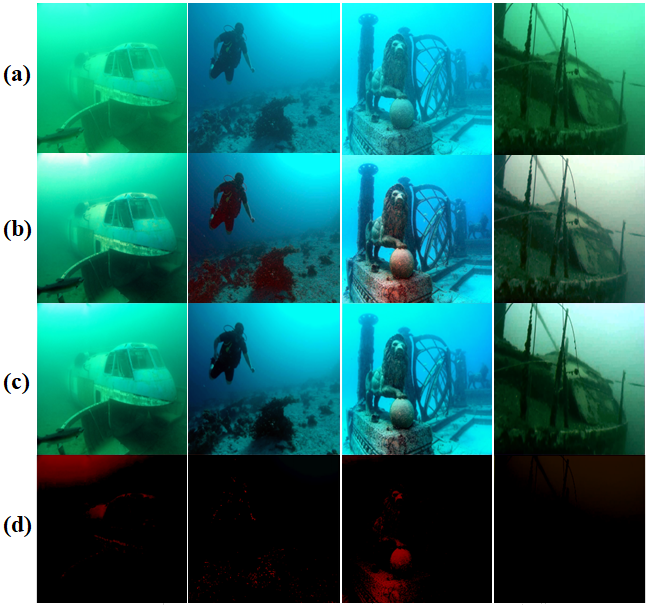}
    \centering
    \caption{Samples of the images presenting unbalance in color channels and outlier pixels treated by the attention module. (a) Input image; (b) Output image; (c) Output image with the attention module inserted in degradation function; (d) Image from the attention module algorithm.}
    \label{fig:outlier3}
\end{figure}

In this section we evaluate the method with the proposed attention module inserted in the pipeline of the model.
Fig. \ref{fig:outlier3}(a) shows samples of images presenting outlier regions (pixels). Fig. \ref{fig:outlier3}(b) shows output images of the method without attention module (baseline) as discussed in the sections above. The outlier regions can be observed in areas presenting high-intensity or color distortion in the images. As an example of this color distortion, the image showing the lion statue highlights an area in red. This area corresponds to an outlier region, generated in the histogram stretching procedure due to the unbalance of the color channels. The red channel in this image presents a narrow dynamic range, which produces a high normalization factor during the histogram stretching. In Fig. \ref{fig:outlier3}(c) are show the output images with attention module present in the pipeline of the model. The high-intensity areas are limited due to the attention module action. It is highlighted that there is a redistribution of the light over a wider area, and a smoothing in visual perception of these regions. In Fig. \ref{fig:outlier3}(d) are showed the images from the algorithm of the attention module. These images are obtained from (\ref{attention}). In the pipeline of the model, the degradation function uses these images to drive the autoencoder to reduce the high-intensity areas. \par

\begin{table}[ht]
  \centering
\small
  \caption{Quantitative results for the model (Mod) and the model with the attention module (Mod+Att). Results for no-reference metrics UIQM, UCIQE, CCF, and NIQE.}
  %\begin{normalsize}
    \label{table:3}
    \begin{tabular}{rcc}\hline{\Large \strut}
             & Mod &  Mod+Att \\ \hline
        UIQM $\uparrow$ & 2.649 & 2.483 \\
        UCIQE $\uparrow$ & 4.623 & 5.050 \\
        CCF $\uparrow$ & 23.49 & 22.02 \\
        NIQE $\downarrow$ & 4.415 & 4.750 \\ \hline
    \end{tabular}
  %\end{normalsize}
\end{table}

We evaluated the model with the attention module added to the pipeline, and compared it with the baseline model (without the attention module). Table \ref{table:3} indicates the results for no-reference metrics. The performance of the model with attention module is similar to the baseline model. It shows slightly lower results for the UIQM, CCF, and NIQE metrics, and a higher score for the UCIQE metric. Similarly to the previous section, the performance related to the full-reference metrics was performed using the images of Turbid/Milk dataset, and the resulting scores are indicated in Table \ref{table:4}. The model with attention model shows performance is very close to the baseline model, and is in agreement with the results for no-reference metrics. In both cases, the differences in performance reflect the action of the attention module.

\begin{table}[ht]
  \centering
%\small
  \caption{Quantitative results for the model (Mod) and the model with the attention module (Mod+Att) for full-reference metrics: PSNR, MSE, SSIM, CIEDE2000 (CIE2K), FSIM, FSIMc, and GSMD. The adopted reference image is the expanded \textit{dynamic range}, and white balanced version of the image 'ref' from Milk dataset, indicated in the Fig.  \ref{fig:ref_turbid}(c).}
  %\begin{normalsize}
    \label{table:4}
    \begin{tabular}{rcc}\hline{\Large \strut}
             & Mod &  Mod+Att \\ \hline
        PSNR $\uparrow$ & 17.65 & 17.24 \\ %\hline
        MSE $\downarrow$ & 0.023 & 0.027 \\ %\hline
        SSIM $\uparrow$ & 0.706  & 0.695 \\ %\hline
        CIE2K $\downarrow$ & 15.32 & 16.83 \\ %\hline
        FSIM $\uparrow$ & 0.979 & 0.978 \\ %\hline
        FSIMc $\uparrow$ & 0.763 & 0.762 \\ %\hline
        GSMD $\downarrow$ & 0.023 & 0.023 \\ \hline
    \end{tabular}
  %\end{normalsize}
\end{table}

\section{Conclusion}

In this work, we presented an image enhancement proposal based on a deep learning approach. Our method uses real UW images only. The algorithm employs an image description, which is inspired by the IFM, but is color space contextualized. Our algorithm degrades properly the output image of the autoencoder, and replaces it in the loss function by the distorted image. This procedure \textit{misleads} the training of the neural network, allowing it to learn how to correct the imposed distortion. The method requires no prior or ground-truth images. Our main assumption states that the natural degradation of an UW image can be synthetically increased more easily than removed from the image. The results showed the effective action of the method, related to color preservation and contrast improvement. Furthermore, we propose an attention mechanism to tackle saturated regions generated in images, presenting unbalance in color channels and strong outlier pixels. A comparative study was performed involving no and full-reference image quality metrics. In both cases, the proposed method provided good performance. However, the results from the no-reference metrics allowed a better assessment of the compared methods, when analyzed jointly. In future work, we intend to extend our methodology to UW applications, as well as different domains that require imagery denoising.

\section*{Acknowledgments}
The authors would like to thank to National Council for Scientific and Technological Development (CNPq) and the Coordination for the Improvement of Higher Education Personnel (CAPES). In addition, the authors are grateful to Dr. Chongyi Li who kindly provided access to the images from the UIEBD database.

%%Vancouver style references.
\bibliographystyle{cag-num-names}
\bibliography{refs}

%\section*{Supplementary Material}

%Supplementary material that may be helpful in the review process should
%be prepared and provided as a separate electronic file. That file can
%then be transformed into PDF format and submitted along with the
%manuscript and graphic files to the appropriate editorial office.

\end{document}